\documentclass{webofc}
\usepackage[varg]{txfonts} 
\usepackage{xcolor}

\newcommand{\blue}{\textcolor{blue} }
\newcommand{\red}{\textcolor{red} }

\begin{document}
\title{Melting and freeze-out conditions of hadrons in a thermal medium}

\author{\firstname{Juan M.} \lastname{Torres-Rincon}\inst{1}\fnsep\thanks{Speaker, \email{torres@fias.uni-frankfurt.de}} \and
        \firstname{Joerg} \lastname{Aichelin}\inst{1,2} \and
        \firstname{Hannah} \lastname{Petersen}\inst{1,3,4} \and
        \firstname{Jean-Bernard} \lastname{Rose}\inst{1,3}  \and
        \firstname{Joseph} \lastname{Tindall}\inst{5}
}

\institute{Frankfurt Institute for Advanced Studies, Ruth-Moufang-Str. 1, 60438 Frankfurt am Main, Germany
\and
           Subatech, UMR 6457, IN2P3/CNRS, Universit\'e de Nantes, \'Ecole de Mines de Nantes, 4 rue Alfred Kastler, 44307 Nantes, France
\and
           Institute for Theoretical Physics, Goethe University, Max-von-Laue-Str. 1, 60438 Frankfurt am Main, Germany
\and
           GSI Helmholtzzentrum f\"ur Schwerionenforschung, Planck-Str. 1, 64291 Darmstadt, Germany
\and
           Department of Physics, University of Bath, Claverton Down, Bath, United Kingdom
          }

          \abstract{We describe two independent frameworks which provide unambiguous de-
terminations of the deconfinement and the decoupling conditions of a relativistic gas at
finite temperature. First, we use the Polyakov-Nambu-Jona--Lasinio model to compute
meson and baryon masses at finite temperature and determine their melting temperature
as a function of their strangeness content. Second, we analyze a simple expanding gas
within a Friedmann-Robertson-Walker metric, which admits a well-defined decoupling
mechanism. We examine the decoupling time as a function of the particle mass and cross
section. We find evidences of an inherent dependence of the hadronization and freeze-out
conditions on flavor, and on mass and cross section, respectively.}

\maketitle

\section{Introduction}
\label{intro}

Statistical thermal models~\cite{thermalmodels} have been very successful in characterizing the thermodynamical 
properties of the hadronic medium created in heavy-ion collisions at the chemical freeze-out. By simply
measuring the total multiplicities of particles, the temperature $T_f$ and the baryochemical potential
$\mu_{B,f}$ can be extracted with a satisfactory statistical confidence. However, the assumption of a single
$T_f$ entails a tension between multistrange and nonstrange baryons in the thermal fits made to the data
of the ALICE collaboration~\cite{alice}. The $\Xi$ and $\Omega$ baryons seems to prefer a larger freeze-out temperature
(with a difference of $\sim$16 MeV) than the $T_f$ for protons.

On the other hand, the computation of generalized susceptibilities by lattice-QCD shows that
the inflection point of the 2nd-order susceptibility---indicator of the crossover transition---is located
for the strange sector at a temperature which is around 15 MeV larger than the one for its light
counterpart~\cite{rene}. Therefore, there exist indications of a flavor hierarchy in (at least) two different
observables: the QCD phase transition and the freeze-out temperatures~\cite{sandeep}. This issue have created
much interest at this SQM2017 conference, e.g.~\cite{sqm}.

In this contribution we provide qualitative evidences of such a hierarchy by presenting two separate studies. The QCD deconfinement temperature will be accessed by the melting temperature of
hadrons in the PNJL model of strong interactions. The freeze-out temperature will be represented by
the decoupling temperature of an expanding system of particles using a relativistic transport model.

\section{Melting temperature hierarchy in the PNJL model~\label{PNJL}}

The PNJL model is an effective realization of the low-energy dynamics of QCD~\cite{njl}. Thanks to its
Lagrangian formulation it can be easily extended to finite densities and temperatures. The dynamical
degrees of freedom are $N_f = 3$ massive quarks. Nevertheless, colorless states like mesons and baryons
can also be addressed in this model with the use of few-body techniques~\cite{benjamin}. Mesons are modeled as
quark-antiquark bound states which are stable in vacuum. At higher temperatures mesons can melt
into pairs. This happens at the so-called Mott (or melting) temperature, which is understood as the
effective deconfinement temperature (as the PNJL model lacks true confinement).

\begin{figure}[ht]
\centering
\includegraphics[width=5.8cm]{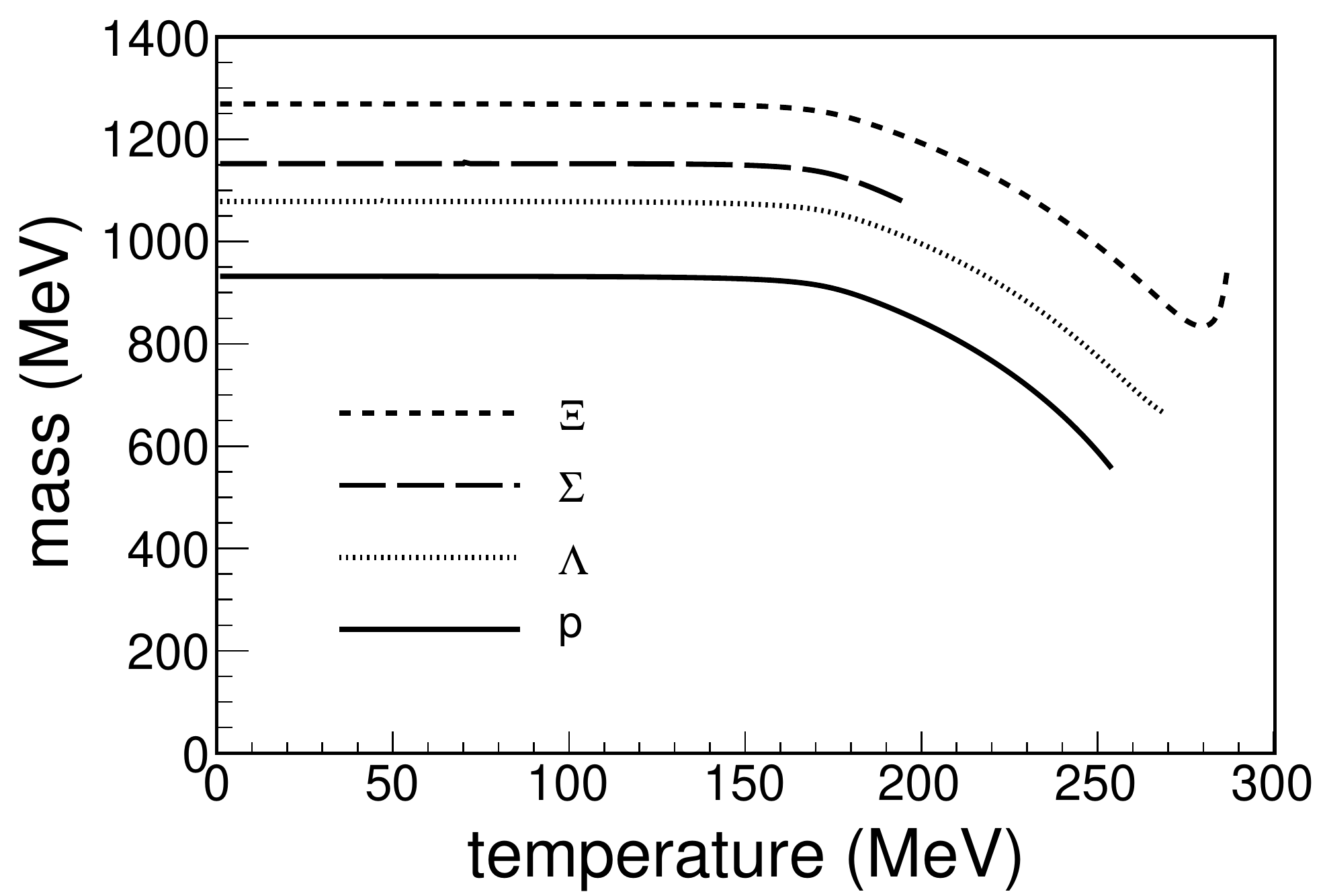}
\includegraphics[width=5.8cm]{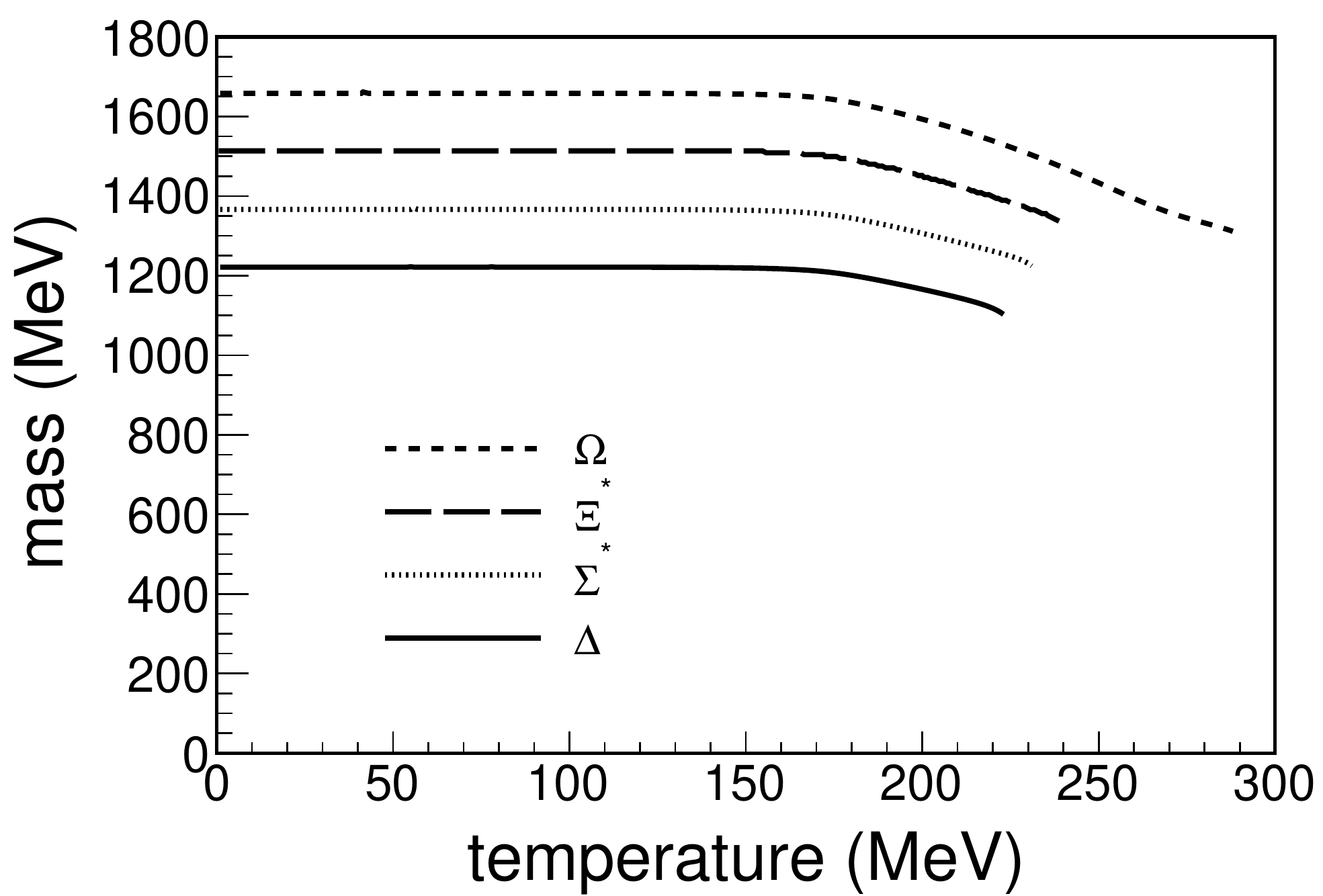}
\caption{Masses of baryons belonging to the flavor octet (left panel) and decuplet (right panel) representations
as functions of the temperature until the melting (or Mott) temperature.}
\label{fig-1}       
\end{figure}

Similar considerations apply to bound states of 2 quarks or diquarks, and to bound states of a
diquark and a quark. The later are identified with physical baryons when they belong to a singlet
color representation. In Ref.~\cite{benjamin} we computed the masses and Mott temperatures of mesons and
baryons in several flavor/spin channels. A summary of the melting temperatures is shown in Table~\ref{tab:njl}.
In addition, the baryons masses versus temperature (up to the Mott temperature) are plotted in Fig.~\ref{fig-1}.

\begin{table}[ht]
\centering
\caption{\label{tab:njl} Mott (or melting) temperatures of several mesons and baryons calculated in the PNJL model~\cite{benjamin}.}
\begin{tabular}{|c|c|c|c|c|c|c|c|c|}
\hline
\bf{Meson} & $\pi$ & $K$ & $\eta$ & $\eta'$ & $\rho$ & $K^*$ & $\omega$ & \red{$\phi$} \\
\hline
$T_{Mott}$ (MeV)& 282 & 286 & 245 & 0 & 253 & 266 & 253 & \red{382} \\
\hline 
\hline
\bf{Baryon} & \red{$p$} & $\Lambda$ & $\Sigma$ & \red{$\Xi$} & $\Delta$ &  $\Sigma^*$ & $\Xi^*$ & \red{$\Omega$} \\
\hline
$T_{Mott}$ (MeV)& \red{254} & 269 & 195 & \red{287} & 223 & 231 & 239 & \red{288} \\
\hline 
\end{tabular}
\end{table}

Several conclusions can be extracted from Table~\ref{tab:njl} and Fig.~\ref{fig-1}: {\bf 1)} The effective deconfinement
temperature is not universal. {\bf 2)} The melting temperatures of $\Xi$ and $\Omega$ are larger (ca. 30 MeV) than
the one for the proton. This relative ordering might be responsible for the situation observed in the
statistical thermal fits [2]. {\bf 3)} Multistrange states generally have larger melting temperatures (except
the $\Xi^*$). In particular, the $T_{Mott}$ of the $\phi$ meson (strange-antistrange pair) is abnormally large.

These results point to a hierarchy in the deconfinement conditions, with a larger temperature for
the strange sector than the light one, as reflected in lattice-QCD results for the susceptibilities~\cite{rene}.

\section{Expanding system and its decoupling}

In a completely different approach we have focused on the freeze-out mechanism and its dependence
on the particle properties. For this case we have used a transport model to describe a system of
relativistic particles interacting via constant cross sections. SMASH (Simulating Many Accelerated
Strongly-interacting Hadrons) is a newly developed transport model that aims to describe the hadronic
stage of heavy-ion collisions with applications from GSI energies to LHC physics~\cite{smash}. We have
adapted SMASH to work with a background Friedmann-Robertson-Walker metric so that the system
exhibits a Hubble-like expansion in time. We use a spherical volume filled with a large number of
particles, initialized with an isotropic, homogeneous spatial distribution.

For a particular initial condition, we were able to reproduce the exact analytical solution of the
massless Boltzmann equation found in Ref.~\cite{bazow}. This represents a nontrivial check for the SMASH
code in a far-from-equilibrium scenario~\cite{tindall}.

For massive particles, we apply a Hubble expansion whose initial rate is much smaller than the
scattering rate. In this regime the system is able to reach and maintain equilibrium during the expan-
sion. When the expansion rate becomes larger than the scattering rate, the equilibrium state cannot be
maintained anymore, as the number of collisions is not sufficient to redistribute the momentum.

\begin{figure}[ht]
\centering
\includegraphics[width=4.5cm]{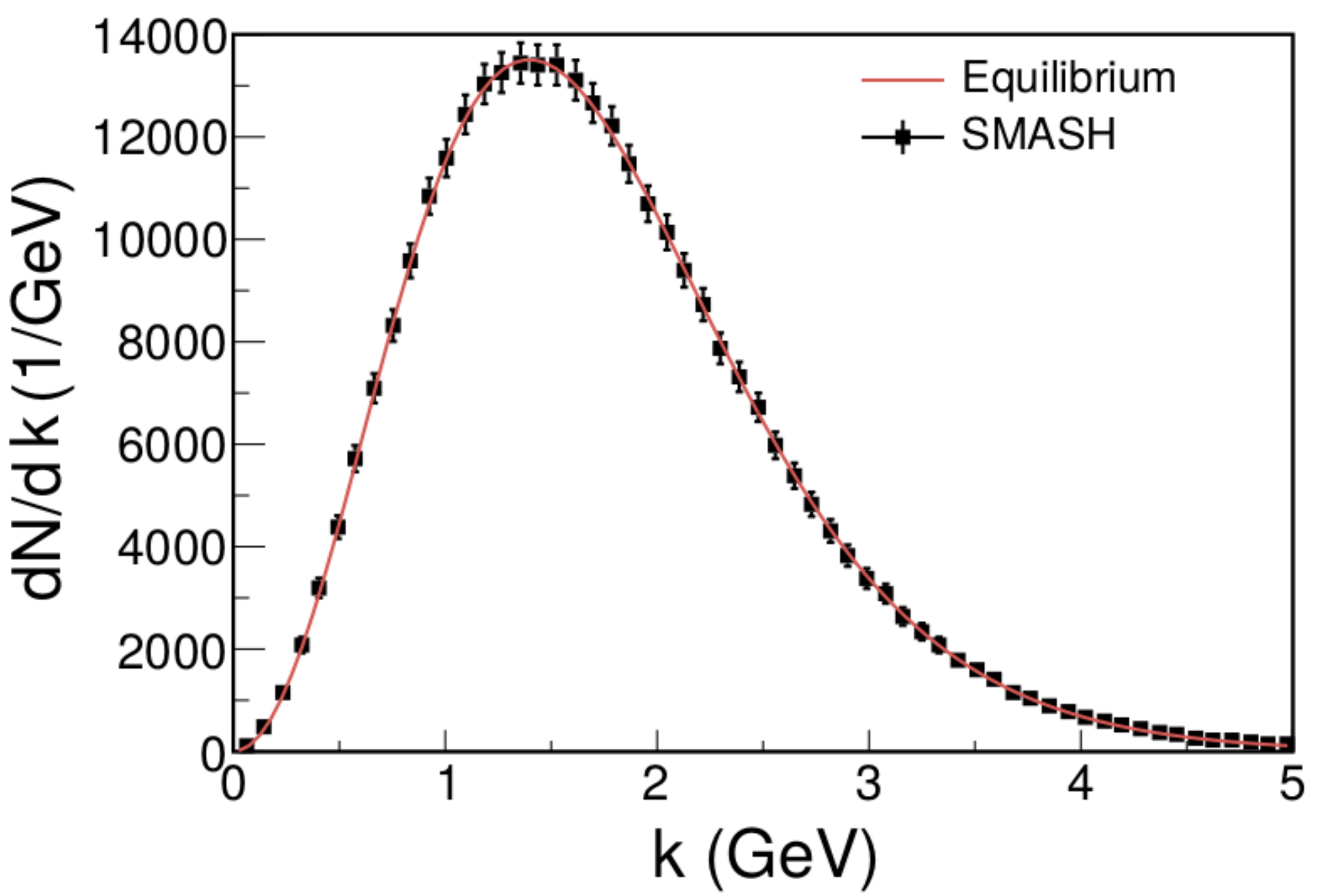}
\includegraphics[width=4.5cm]{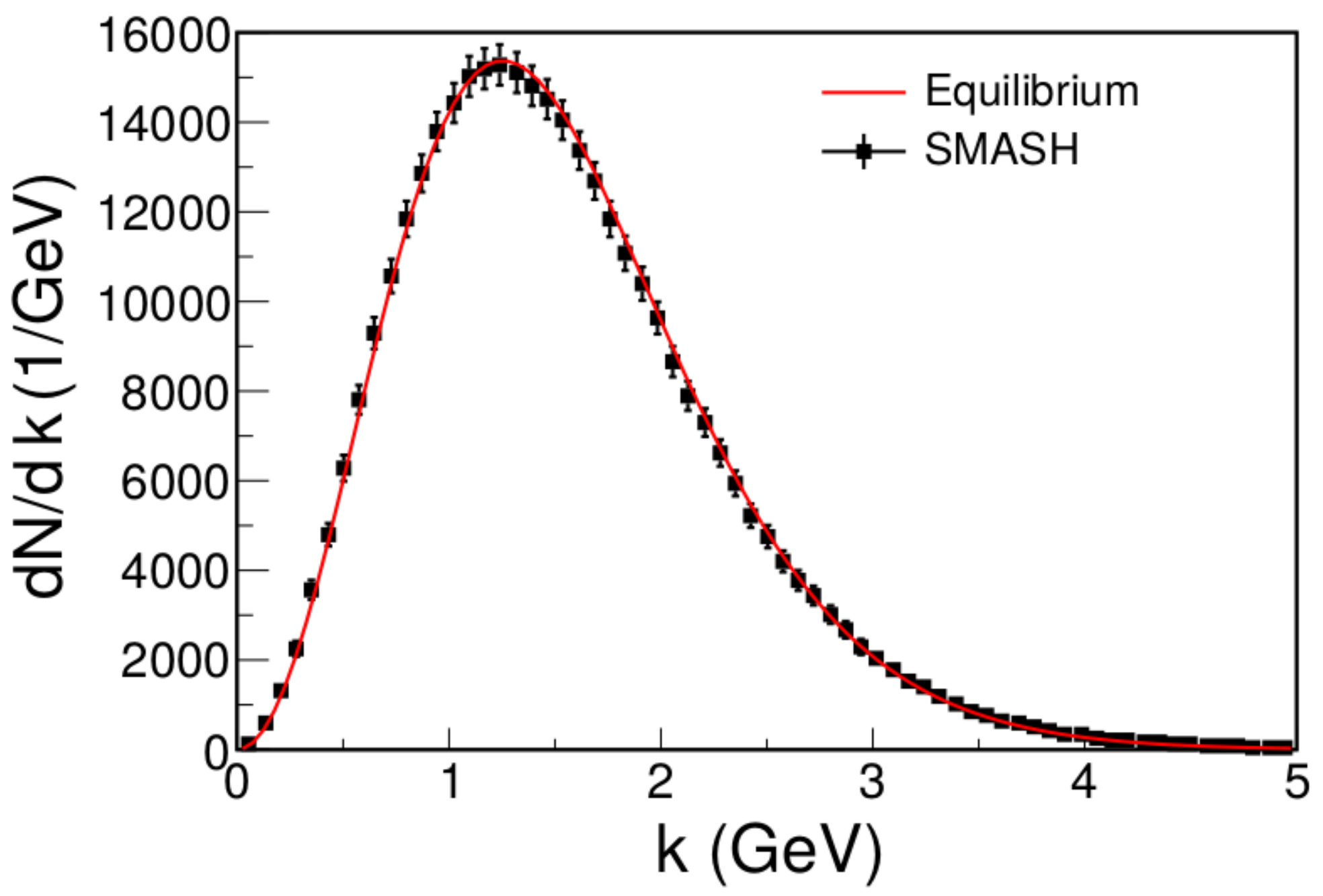}
\includegraphics[width=4.5cm]{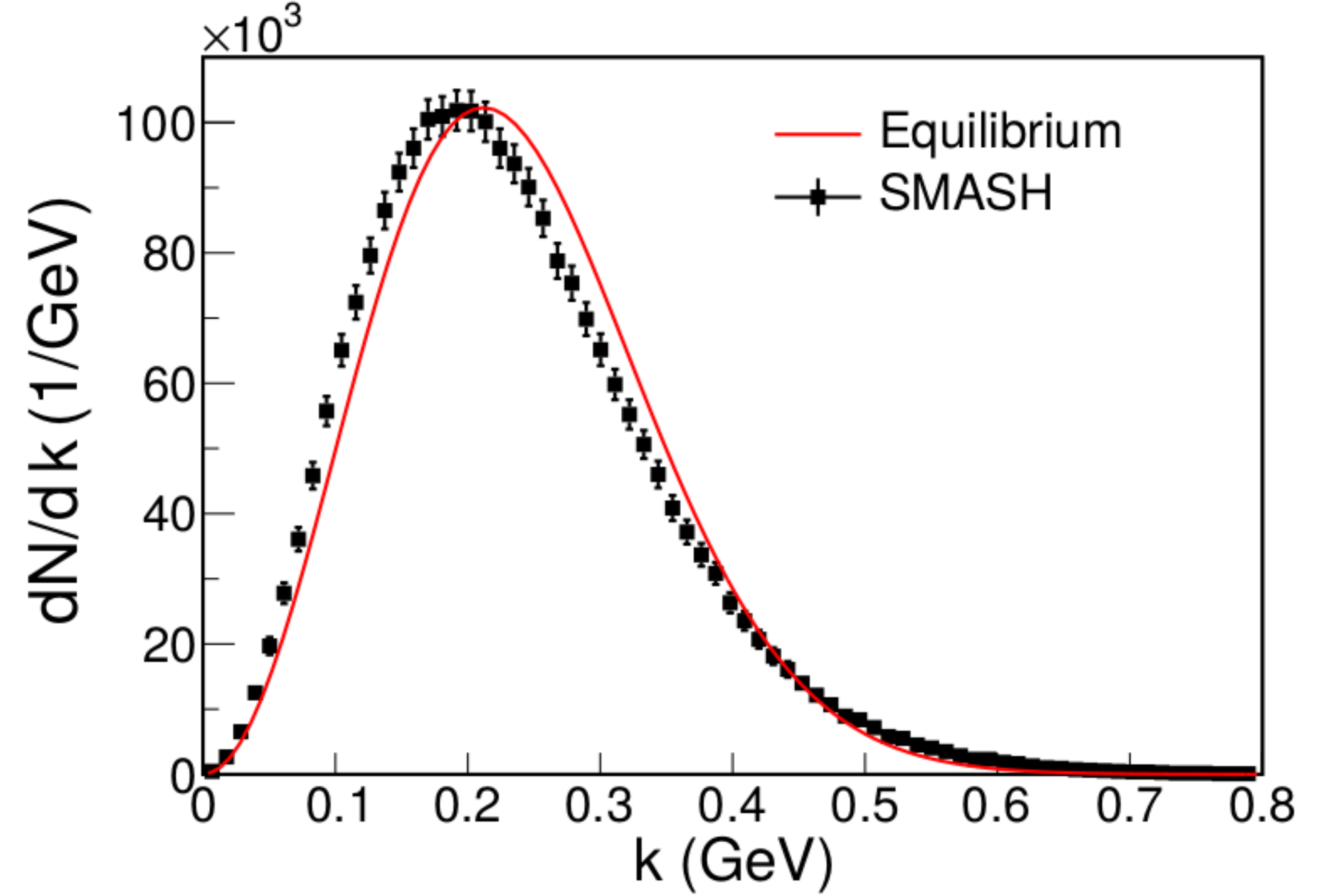}
\caption{Distribution function at different times of the Hubble expansion as a function of the particle momentum.
From left to right $t=0.1, 5$ and $20$ fm. In solid line we show the predicted equilibrium distribution for each time.}
\label{fig:dec1}
\end{figure}

From Fig.~\ref{fig:dec1} we can observe that SMASH perfectly reproduces the theoretical distribution function
(solid line) until decoupling happens soon after the middle panel. At $t=20$ fm the distribution from
SMASH does not match the equilibrium expectations. Taking into account the absence of collisions
and the redshift in momentum, the distribution after the freeze-out (decoupling) reads

\begin{equation} 
f_{freeze-out} (t,k) = f_{equilibrium} \left( t_D,k \red{a(t)/a_D} \right) \propto \exp \left[ - \frac{ \sqrt{ \left( k \red{ a(t) /a_D} \right)^2+m^2} }{\blue{T_D}} \right] \ ,
\end{equation}
where $t_D$ is the decoupling time, $\red{a(t)}$ is the scale factor, and $\red{a_D} , \blue{T_D}$ are the value of the scale factor and the temperature at the decoupling time, respectively.

\begin{figure}[t]
\centering
\includegraphics[width=5.5cm]{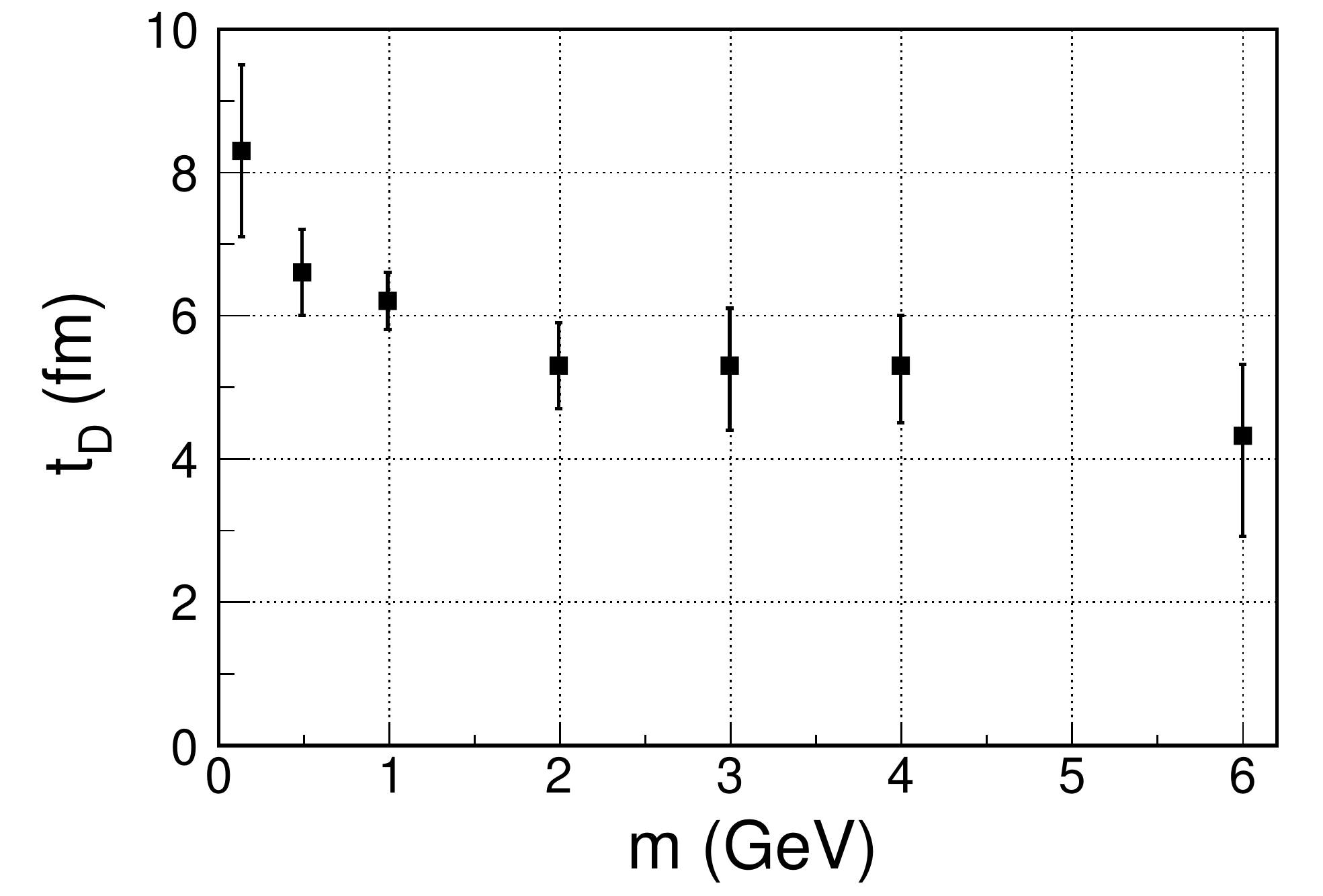}
\includegraphics[width=5.5cm]{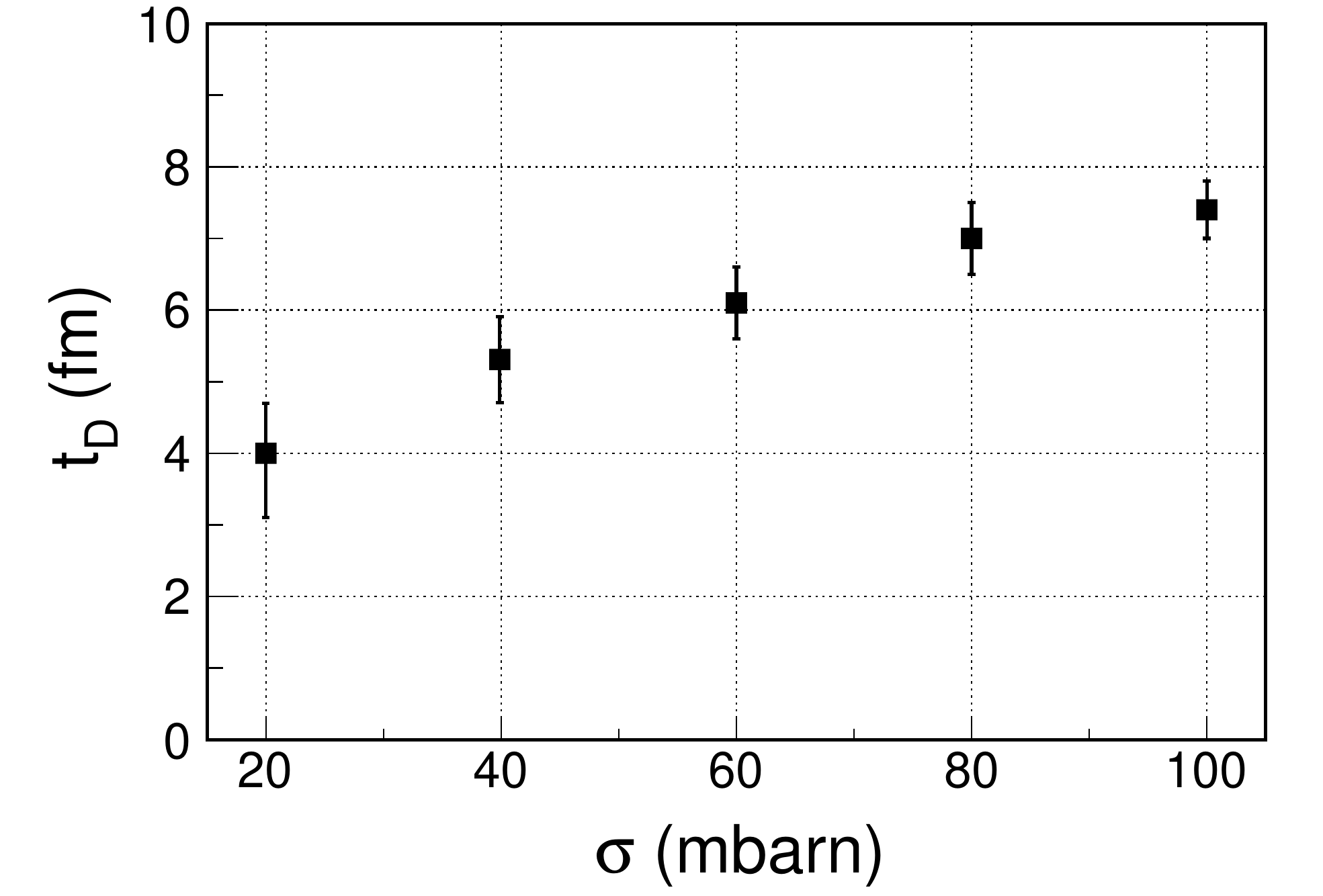}
\caption{Decoupling time versus the mass of the particles with fixed cross section $\sigma=40$ mb (left panel), and
versus their total cross section with fixed mass $m=2$ GeV (right panel).}
\label{fig:dec2}       
\end{figure}

The fit to this function provides the value of the decoupling time and temperature. In the example
of Fig.~\ref{fig:dec1} it gives $t_D = 5.3 \pm 0.6$ fm. Repeating this procedure for particles with different masses and
cross sections, we obtain the results in Fig.~\ref{fig:dec2} for the extracted freeze-out times.

The main conclusions of this study are: {\bf 1)} The more massive particles freeze out earlier (at higher
temperature). {\bf 2)} The larger the cross section is, the later the freeze-out appears (lower temperature).
Two caveats are in order. This scenario is closer to the kinetic freeze-out (the absence of interactions) 
rather than the chemical freeze-out (relevant for statistical thermal fits). So far, inelastic processes are absent in our study. In addition, we have not explicitly introduced strangeness in the system.
However, one can notice that strange hadrons generally entail larger masses, and smaller cross sections
in comparison with their lighter counterparts. According to the results in Fig.~\ref{fig:dec2}, the combined effect
would result in an earlier freeze-out for the hadrons with strangeness (i.e. higher values of $T_f$).

\section{Conclusion}

We provide evidences of a flavor-nonuniversal deconfinement and freeze-out conditions that support
previous results in this directions~\cite{sqm,sandeep,rene}. On one hand, using the PNJL model we observe that multistrange hadrons melt at higher temperatures rather systematically. This points towards an earlier
confinement transition for the strange sector in heavy-ion collisions. On the other hand, the decoupling mechanism in an expanding spacetime provides direct indications that more massive and less
interacting particles freeze out earlier and at higher temperatures. This observation supports the higher
freeze-out temperature found for (multi)strange hadrons in statistical thermal fits~\cite{alice}.

 \vspace{2mm}

\begin{acknowledgement}
Work supported by DAAD (Germany), Helmholtz Young Investigator Group VH-NG-822 (Helmholtz Association and GSI), the HIC for FAIR within the framework of the LOEWE program (Hesse, Germany), and
Project No.FPA2013-43425-P (Spain), and Program TOGETHER from R\'egion Pays de la Loire and the I3-Hadron Physics (EU). We acknowledge LOEWE-Center for Scientific Computing for computing resources.
\end{acknowledgement}

\end{document}